# Metal to insulator transition, colossal Seebeck coefficient and ultralow thermal conductivity in solution-processed monodispersed nickel nanoparticles

Vikash Sharma[1], Gunadhor Singh Okram[1] and Yung-Kang Kuo[2]

[1]UGC-DAE Consortium for Scientific Research, University Campus, Khandwa Road, Indore 452001, Madhya Pradesh, India.
[2]Department of Physics, National Dong-Hwa University, Hualien 97401, Taiwan
okram@csr.res.in and vikash@csr.res.in

**Abstract** We report here metal to insulator transition, colossal Seebeck coefficient and ultralow thermal conductivity κ ($\frac{1}{175}$th of its bulk value, significantly smaller than many well-known thermoelectric (TE) materials and silicon, showing potential applications in TEs, electronics and photonics for heat dissipation) in monodispersed well characterized Ni nanoparticles (NPs). As a consequence, TE power factor and figure of merit (ZT) are significantly enhanced compared to their bulk counterpart. Interestingly, a systematic crossover from metallic to semiconducting to finally electrically insulating behavior, large negative temperature coefficient of resistance (TCR) and n-type conduction to p-type conduction with decrease in particle size have been observed. These results are mainly attributed to formation of metal/organic interfaces, enhancement in local electronic density of sates (DOS) and multiscale electron and phonon scattering by various defects. Thus, this study will open a new avenue to make better TEs through incorporation of such NPs in semiconducting hosts.

## 1. Introduction

TE materials provide an efficient solution to utilize waste heat with many advantages such as cost-effectiveness, free from moving parts, rapid response time and small size[1–10] since they are capable of converting heat directly into electricity. Their conversion efficiency strongly rely on dimensionless figure of merit, ZT = $\frac{S^2\sigma}{\kappa}$T, where S is the Seebeck coefficient or thermopower, $\sigma$ is electrical conductivity, $\kappa = \kappa_e + \kappa_l$ is thermal conductivity, wherein $\kappa_e$ and $\kappa_l$ are electronic and lattice part of κ, and T is the absolute temperature[11]. Consequently, competitive TE material prefer an interactive combination of high power output that rely on high power factor ($S^2\sigma$) and low κ to sustain the large temperature gradient even at high T[12]. Therefore, these transport parameters make TE materials extremely difficult to improve their ZT in ordinary materials. Fortunately, there are two main approaches to enhance TE performance: first is to enrich the phonon scattering by incorporating defects, nanoinclusions and blending of nanocrystals or nanocomposites, and second is to improve electrical transport by band structure engineering including resonant doping, band flattening and band convergence[1,13]. Also, making the complex materials by means of size and morphology-controlled hybrid inorganic/organic material or nanocomposites considerably enhance the TE properties[3,14,15].

Accordingly, manipulating materials with improved TE properties is a key challenge for new generation of TEs. In this context, colloidal NPs synthesized using bottom-up approach have been overlooked compared to other established methods although it can precisely be tuned their electronic properties by controlling the particle size, size distribution and shape via reaction conditions and variety of surfactant/s. These NPs may provide nearly precisely controlled platform for exploring the interesting physical phenomena such as metal to insulator transition (MIT)[16], superparamagnetic behaviour of ferromagnetic material[17] and ferromagnetism in antiferromagnetic materials[18]. Particularly, reduction in particle size reduces κ and $\sigma$, though at different levels, to the advantage of a TE. They are due to increase in various defects such as grain boundaries (GBs), point defects, dislocations, surfactant matrix and quantum size effect (QSE), which in turn enhance S mainly due to modified electronic structure. The QSE in metal NPs is more dominant as the size decreases since the number of electrons (N) to be filled up to its bulk Fermi energy ($E_F$ ~ 5 eV) tends to be very limited (N ~ $10^3$) compared to its bulk counterpart wherein energy levels are nearly continuous as the electron density of ~$10^{22}$ per cm$^{-3}$ is very high. Consequently, energy levels become discrete in QSE metal particle and level spacing, called Kubo gap ($\delta = 3E_F/4N$)[19,20,21], becomes much broader. Additionally, there is also the probable transition from metallic to non-metallic state when $\delta > k_B T$ and Fermi level lies between the levels[22]. Metallic NPs have aroused extensive attention in TEs as they are extremely effective to reduce κ and enhances the electrical transport of the host semiconductor after incorporating them[2].

However, adsorbed molecules of surfactant may also perturb the electronic density of states (DOS) of metal NPs, and there may be possibility of charge transfer between metal and organic semiconductor. Thus, these surface-bound molecules on NPs' surface reduce the electrical properties owing to decrease in mobility of charge carriers that is mainly due to enhanced scattering, high interfacial densities and impurities[23]. The interface between metal and organic semiconductor may significantly modify the electrical transport and thermoelectric properties of NPs[24]. Furthermore, there is dominantly increasing coverage or density/thickness of surfactant as a barrier for transport of electrons as the size decreases[25]. Therefore, the transport of heat and electricity are mediated by the density and chemistry of the metal/organic material interfaces, and the volume fractions of nanocrystal cores and surface ligands[14]. These studies carried out in monodispersed Ni NPs have been presented here.

## 2. Experimental
### 2.1 Sample preparation

Nickel acetylacetonate, Ni(acac)$_2$ (95%), TOP (90%) and OA (70%), purchased from Sigma Aldrich, were used as received. Typically, 3 g Ni(acac)$_2$ and 10 ml OA was mixed in three-neck round bottom flask and heated at 210 °C for 2 hours under a nitrogen atmosphere. The reaction product was cooled down to room temperature and centrifuged after the addition of n-hexane and ethanol to extract the NPs. This was done three times to remove excess OA, TOP or acetate and then the particles were dried at 60 °C for characterizations. This sample was coded as Ni1. To tune the particle size, we followed the procedure described earlier[17,25–27]. 0.25 ml of preheated TOP at 200 °C is added in a solution of 3 g Ni(acac)$_2$ in 10 ml OA, already degassed at 120 °C for 30 mins with remaining reaction conditions the same. This sample was coded as Ni2. Similarly, samples coded as Ni3, Ni4, Ni5, Ni6 and Ni7 were prepared with TOP 1 ml, 3 ml, 5 ml, 7 ml and 10 ml, respectively. Details of sample preparation conditions are summarized in table 1.

**Table 1** Sample preparation conditions, crystallite size, TEM size and lattice parameters.

| Sample | OA (ml) | TOP (ml) | Crystallite Size (nm) | TEM size (nm) | Lattice parameter (Å) |
|---|---|---|---|---|---|
| **Ni1** | 10 | Nil | 23.1±0.3 | 70.5±4.2 | 3.5325 |
| **Ni2** | 10 | 0.5 | 15.3±0.2 | - | 3.5314 |
| **Ni3** | 10 | 1 | 10.9±0.4 | - | 3.5295 |
| **Ni4** | 10 | 3 | 8.4±0.2 | 10.9±0.5 | 3.5281 |
| **Ni5** | 10 | 5 | 6.8±0.3 | - | 3.5275 |
| **Ni6 fcc** | 10 | 7 | 3.2±0.2 | - | 3.5277 |
| **hcp** | | | - | - | a=b=2.6554, c=4.3599 |
| **Ni7 fcc** | Nil | 10 | 1.3±0.3 | 4.5±0.1 | 3.5232 |
| **hcp** | | | - | - | a=b=2.6249, c=4.3251 |

### 2.2 Characterizations

Bruker D8 Advance X-ray diffractometer with Cu Kα radiation (0.154 nm) for laboratory and beamline BL-18, KEK, Japan for synchrotron X-ray diffraction (XRD) measurements of Ni NPs were performed on powder samples. TEM measurements were performed using TECHNAI-20-G² on NPs dispersed over carbon-coated TEM grids by drop-casting the well-sonicated NPs. Resistance measurements were performed on macroscopic cold-pressed samples using four-point probes of Ni1 to Ni5 and two-point probes of Ni6 and Ni7, and Seebeck coefficient measurements using differential direct current setup in the temperature range of 5–300 K in a specially designed commercially available Dewar[28,29]. Thermal conductivity was measured using a dc pulse laser technique in the temperature range of 10 - 300 K[30].

## 3 Results and discussion
### 3.1 X-ray diffraction (XRD) and transmission electron microscopy (TEM)

The laboratory X-ray source XRD patterns of Ni1 to Ni7 are depicted in figure 1 (a). The three peaks appeared in Ni1 to Ni6 clearly confirm the face-centred cubic (fcc) crystal structure of Ni, without any impurity peak. All the three peaks of Ni7 are not visible, which is clearly seen in synchrotron beamline XRD (figure 1 (a), left inset). The synchrotron radiation XRD of Ni1, Ni2 and Ni6 are consistent with laboratory source XRD. Notably, XRD peaks are broadened with increase in TOP, which is indicative of decrease in crystallite size that is line with earlier reports[17,25–27]. The average crystallite size evaluated from Scherrer formula for Ni1, Ni2, Ni3, Ni4, Ni5, Ni6 and Ni7 is 23.1±0.3, 15.3±0.2, 10.9±0.4, 8.4±0.2, 6.8±0.3, 3.2±0.2 and 1.3±0.3 nm, respectively (table 1). They show systematic decrease in crystallite size as TOP concentration increases.

Rietveld refinements using Fullprof suite software of XRD data are done to calculate the lattice parameters (figure 2 and table 1). It can clearly be seen that as crystallite size decreases down to 6.8 nm, there is a systematic decrease in lattice parameter, and below it, it increases in 3.2 nm NPs, and then again steadily decreased in 1.3 nm NPs (figure 1 (a), right inset) The anomaly in lattice parameter in 3.2 nm NPs suggests that there is size-induced structural phase transition from purely fcc to mixed fcc-hcp (hexagonal closed-packed) structures below ~ 6 nm (table 1), which is consistent with previous report on such NPs[17].

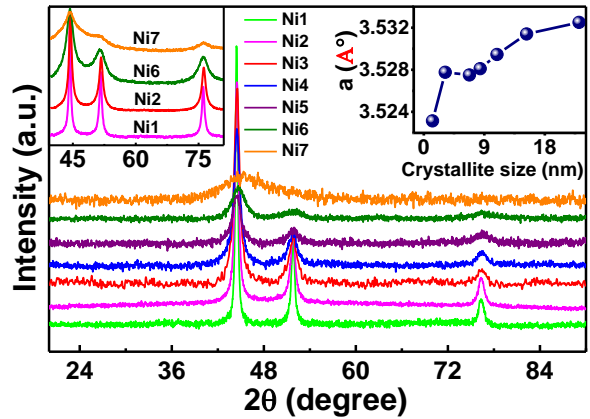

**Figure 1** Laboratory source X-ray diffraction patterns of Ni1 to Ni7. Inset: (left) synchrotron radiation x-ray diffraction of Ni1, Ni2, Ni6 and Ni7 and (right) lattice parameter as a function of particle size.

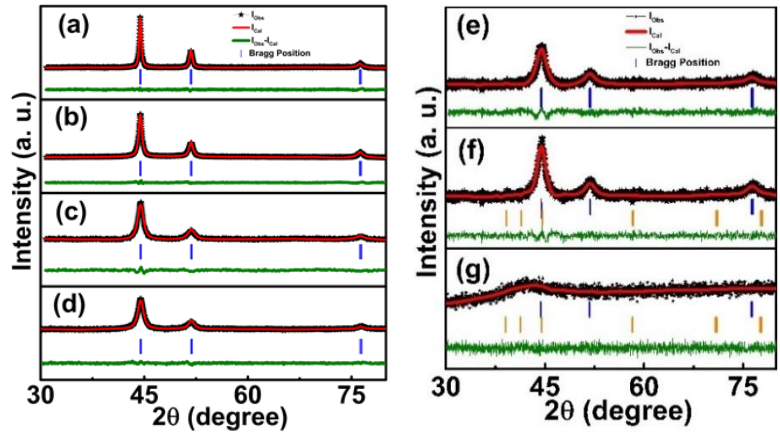

**Figure 2** Rietveld refined X-ray diffraction patterns of (a) Ni1, (b) Ni2, (c) Ni3, (d) Ni4, (e) Ni5, (f) Ni6 and (g) Ni7.

TEM images of Ni1, Ni4 and Ni7 as representatives are shown in figure 3 (a, b, c). The random morphology and broad size distribution of Ni1 can be seen (figure 3 (a) & (d), inset), and Ni4 has narrower size distribution than Ni1 (figure 3 (b) & (e), inset) while Ni7 exhibits reasonably monodispersed NPs with approximately similar morphology (figure 3 (c) & (f), inset). The average particle size is 70.5±4.2, 10.9±0.5 and 4.5±0.1 nm for Ni1, Ni4 and Ni7, respectively. It can be seen that as TOP increases, particle size decreases and size distribution becomes narrower, which is consistent with Scherrer size. Interestingly, increase in broadening with increase in TOP corroborates the decrease in degree of crystallinity with decrease in particle size obtained from selected area diffraction patterns (SAED) (figure 3 (d, e, f), insets). However, we must stress here that Ni3-Ni7 are believed to be associated with nanolattices due to their monodispersity and nanolattice formable capacity of TOP[26].

### 3.2 Electrical resistivity

Figure 4 (a) shows the variation in $\rho$ with T for Ni1 to Ni7 samples. It decreases with decrease in T for Ni1 and Ni2, revealing their metallic behaviour (figure 4 (a), inset), which suggests that scattering of electrons with phonons dominates over other scattering mechanisms[31]. Ni3, Ni4 and Ni5 show metallic behaviour down to near 70 K, 100 K and 170 K, respectively, and below this, $\rho$ starts to increase with decrease in T revealing their semiconducting behaviour (figure 4 (a)); that is, metal to semiconductor (insulator) transition (MIT) prevails. Thus, a temperature-driven MIT in Ni3 to Ni5 is observed, which shifts to higher T with decrease in particle size. Such type of MIT in Au, Ag NPs and Zn nanowire composites have been reported previously[16,32–36], but not reported for Ni NPs so far, to the best of our knowledge. The MIT is completely disappeared in Ni6 that shows semiconducting nature in the whole range of 10 to 300 K. Furthermore, highly enhanced resistivity to nearly insulating behaviour is seen in Ni7, that is fifteen orders of magnitude approximately exponentially from metallic to highly semiconducting nature as particle size reduces (figure 4 (a)).

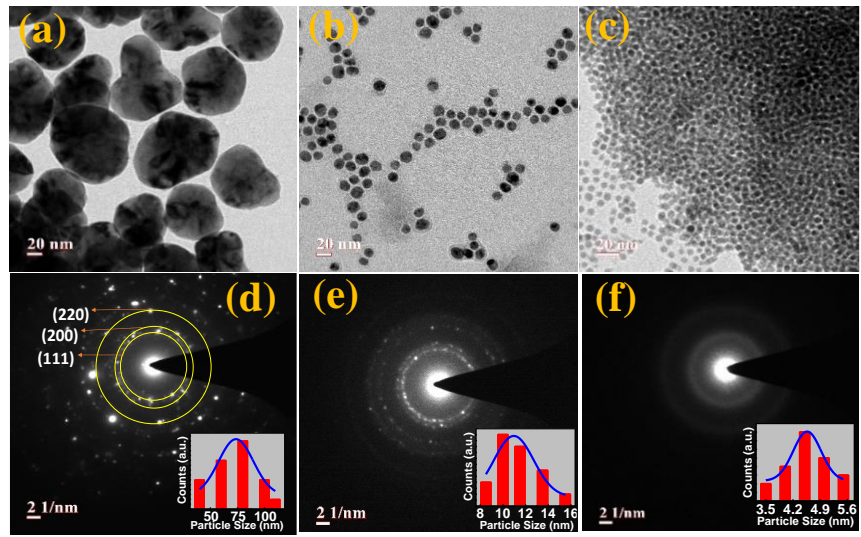

**Figure 3** Transimission electron microscopy images (a, b, c) and selected area electron diffrcation patterns (d, e, f) of Ni1, Ni4 and Ni7, resepctively, and insets (d, e, f) represent their size distributions.

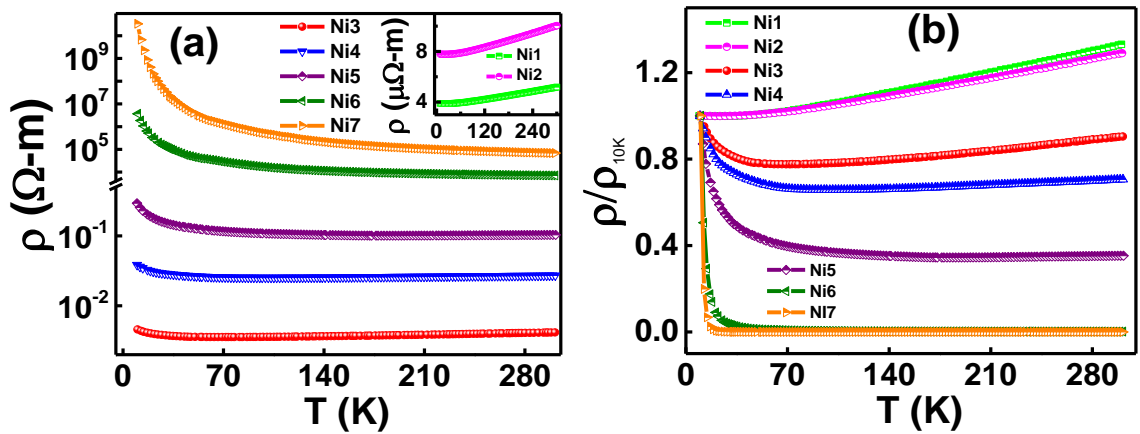

**Figure 4** (a) Electrical resistivity of Ni3 to Ni7; inset: resistivity of Ni1 and Ni2, and (b) variation in ratio of resistivity to resistivity at 10 K ($\rho/\rho_{10K}$).

The ratio $\rho/\rho_{10K}$ decreases with T as well as with decrease in particle size in whole T range (figure 4 (b)), where $\rho_{10K}$ is resistivity at 10 K. This clearly shows that residual resistance ratios (RRR), defined as RRR = $\rho_{300K}/\rho_{10K}$, decrease with decrease in particle size (figure 5 (a)). It indicates worsening quality of the samples as particle size decreases. It is interesting to assess how the particle size influences the resistivity. It shows a logarithmic rise of the order of fifteen as size decreases when we consider it at 300 K and 10 K separately (figure 5 (b)). Temperature coefficient of resistivity (TCR= $\frac{d\rho}{\rho dT}$) is positive in T range 10 K to 300 K for Ni1 to Ni2 (figure 5 (c)). It turns however to negative below near 70 K, 100 K and 170 K in Ni3, Ni4 and Ni5, respectively, and in whole T range in Ni6 and Ni7 (figure 5 (c)). The absolute value of TCR systematically increases with decrease in particle size (figure 5 (d)), and attains a large negative value of around -1.09 K$^{-1}$ and a low -0.0034 K$^{-1}$ at 10 K and 300 K, respectively in Ni7. These trends are attributed to enhanced GBs, point defects, dislocations, surfactant matrix and QSE as particle size decreases.

The MIT in metal NPs assemblies have also been explained in the frame work of Mott-Hubbard and Anderson localization[15,16,33–36]. Electronic transport in NP arrays highly depends upon surface-bound molecules, interparticle distance, and other defects that affects the tunnelling or transmission of electrons between nearest-neighbour NPs[24,15]. The tunnelling rate is inversely related to interparticle distance and height of barrier energy, i.e. lower coupling between NPs leads to decrease in electrical conductivity. In this regard, barrier height for electron transport increases with increase in TOP and simultaneously decreases the coupling between adjacent NPs due to the insulating nature of organic ligands (OA and TOP)[15]. In other words, tunnelling or transmission probability of electrons is decreased with decrease in particle size. As a consequence, semiconducting/ insulating behaviour of the particles themselves as well as the enhanced coverage of the semiconducting surfactant molecules are observed in NPs earlier[15,16,33–36] and presently (figure 5 (b)) with enormous increment in ρ with decrease in crystallite size. Also, the decrease in particle size increases the level spacing or may create the energy gap (or Coulomb gap) at the Fermi level in low temperature, and hence energy dependence on electron relaxation time and concentration of electrons will be significantly different.

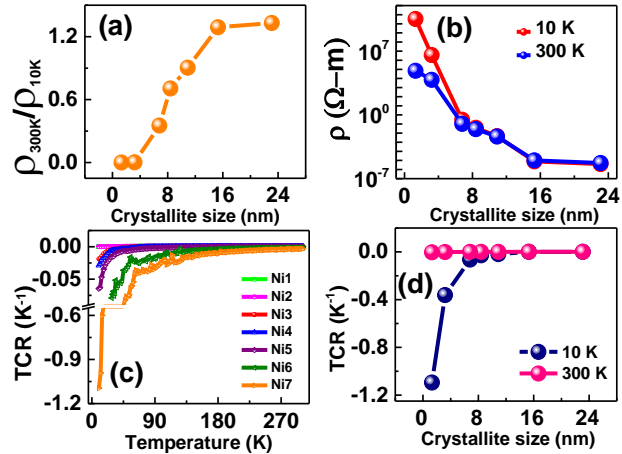

**Figure 5** (a) $\rho_{300K}/\rho_{10K}$ and (b) resistivity at 10 K and 300 K versus crystallite size, (c) Temperature coefficient of resistivity (TCR) for Ni1 to Ni7 and (d) TCR at 10 K and 300 K versus crystallite size.

The metallic behaviour of Ni1 and Ni2 may be mainly associated with electrons being able to conduct through one electronic state to the other inside single NPs as well as easily pass through the GBs and other defects, wherein electron-phonon scattering is major resistance. The T-driven MIT in Ni3, Ni4 and Ni5 would reveal that there is co-tunnelling that resembles variable range hopping (VRH) in these NPs rather than the long-range hopping[33,34]. As the T decreases, electrons can hop via quantum mechanical tunnelling across the GBs that gives rise to semiconducting behaviour below these T's. This transition temperature increases with decrease in particle size clearly indicating that increase in overall barrier height for electrons transport and hence their hop distance decreases as particle size decreases[16]. Semiconducting/ and insulating behaviour of Ni6 and Ni7 in whole T range can thus be further evident. Empirically, when the particle size is larger, the metallic core dominates over the semiconducting surfactant layer (SSL) on the surface of the NPs. As the size decreases, metallic core gradually decreases at the cost of increasing SSL shell. In the medium range (Ni3, Ni4 and Ni5) where MIT exists, this competition goes on until SSL along with QSE-related semiconducting behaviour would overtake completely the metallic core and hence the full-fledged semiconducting nature in Ni6 and Ni7.

In order to check the nature of electron conduction more precisely by means of tunnelling or activated process, semiconducting regime is fitted in the frame work of Efros-Shklovskii-VRH (ES-VRH), Mott-VRH and Arrhenius activation models in low and high T ranges, respectively[16]. In fact, the electronic transport of granular metals wherein some insulating matrix covers the metallic granules and arrays of metallic NPs is mainly driven by phonon-assisted tunnelling or hopping through insulating matrix which separates the NPs. In dense NP arrays, electrons can transport through nearest-neighbour hopping (NNH) and through co-tunnelling between adjacent NPs[37]. The NNH dominates at relatively high temperature while hopping or co-tunnelling dominates at low temperature. Arrhenius behaviour at relatively high temperature can arise from single-electron charging of NPs that predicts

$$\rho \sim \exp(E_c/k_B T) \quad (1)$$

dependence, where $E_c$ is charging or activation energy of a NP and $k_B$ is the Boltzmann constant. The ES-VRH at low temperature accommodates strong Coulomb interactions that predicts

$$\rho \sim \exp(T_{ES}/T)^{0.5} \quad (2)$$

dependence, where $T_{ES}$ is the ES-VRH characteristic temperature which depends upon the dielectric constant of the material and localization length ($\xi$)[16]. This is the extension of VRH of Mott in 3D given by

$$\rho \sim \exp(T_M/T)^{0.25}, \quad (3)$$

where $T_M$ is the Mott characteristic temperature that depends on the electronic DOS at the Fermi level N(E) and $\xi$. The Mott characteristic temperature can be given as

$$T_M = \frac{18}{\xi^3 N(E_F) k_B}. \quad (4)$$

Mean hopping energy, i.e. Mott hopping energy, can be calculated using

$$E_M = \frac{1}{4} k_B T \left(\frac{T_M}{T}\right)^{\frac{1}{4}}. \quad (5)$$

Mott-VRH model is applicable when the average hopping distance ($R_M$) is larger than $\xi$ and separation between nearest-neighbor impurity i.e.

$$\frac{R_M}{\xi} = \frac{3}{8}\left(\frac{T_M}{T}\right)^{\frac{1}{4}} > 1. \quad (6)$$

Notably, in the Mott theory, $N(E_F)$ is taken as constant without electron-electron correlations or Coulomb interactions between charge carriers. Efros and Shklovskii[16] found that at sufficiently low temperature, $N(E)$ cannot be treated as constant due to Coulomb interactions. In granular metals, rich DOS at Fermi level limits the hopping conduction to nearest-neighbours at low temperature. Consequently, Coulomb interactions opens a gap, called Coulomb gap, in the electrical transport. Now,

$$T_{ES} = \frac{2.8e^2}{\varepsilon \xi}. \quad (7)$$

Hopping energy between different sites or ES-VRH hopping energy can be given by

$$E_{ES} = \frac{1}{2} k_B T \left(\frac{T_{ES}}{T}\right)^{\frac{1}{2}}. \quad (8)$$

ES-VRH model is valid when the average hopping distance ($R_{ES}$) is greater than $\xi$ i.e. the ratio of $R_{ES}$ and $\xi$ must be larger than unity i.e.

$$\frac{R_{ES}}{\xi} = \frac{1}{4}\left(\frac{T_{ES}}{T}\right)^{\frac{1}{2}} > 1 \quad (9)$$

Therefore, at sufficiently low T, the electronic transport can be described by ES-VRH that considers Coulomb interaction which vanishes above a certain T. Above this temperature, conductivity can be varied according to Mott-VRH. This crossover from Mott-VRH to ES-VRH with decreasing temperature with opening of Coulomb gap ($\Delta_{CG}$) at the Fermi level is given by

$$\Delta_{CG} \approx \frac{e^3 N(E)^{\frac{1}{2}}}{\varepsilon^{\frac{3}{2}}} \approx k_B \left(\frac{T_{ES}^3}{T_M}\right)^{\frac{1}{2}}. \quad (10)$$

Eq. (3.10) indicates that larger the $N(E)$, wider the $\Delta_{CG}$. The crossover temperature ($T_C$) between Mott-VRH to ES-VRH is given by

$$T_C = 16 \frac{T_{ES}^2}{T_M}. \quad (11)$$

Now, semiconducting regimes of these NPs are first fitted using simple activation or Arrhenius model (eq. 1). However, the fitting quality for the whole semiconducting temperature range is poor (not shown here). This suggests that transport processes cannot be described using a single activation behaviour. But, it gives a better fit at a specific high T regime in each sample. Thus, we fit resistivity using ES-VRH (eq. 2), Mott-VRH (eq. 3) and Arrhenius (eq. 1) models in different temperature regimes (figure 6). Obtained parameters from the fitting and calculation from the above equations (1-11) are listed in tables 2-4. The ES-VRH fits ρ of Ni3 between 10 K-24 K well, Mott-VRH model fits well between 17 K-42 K and simple activation behaviour fits well between 27 K-47 K. This suggests that Mott-VRH and Arrhenius mechanisms coexist in 27 K – 42 K. Clear deviations from ES-VRH above 25 K can be seen for Ni4 (figure 6 (b)), and above this, up to 44 K, it is better-fitted with Mott-VRH. Above 45 K and below the metallic regime, Arrhenius model well-fits it. This crossover from Mott-VRH to ES-VRH manifests opening of the Coulomb gap at the Fermi level. Similarly, the ρ of other samples is fitted using different temperature ranges in which these conduction mechanisms are dominated (see figure 6 & tables 2- 4 for details).

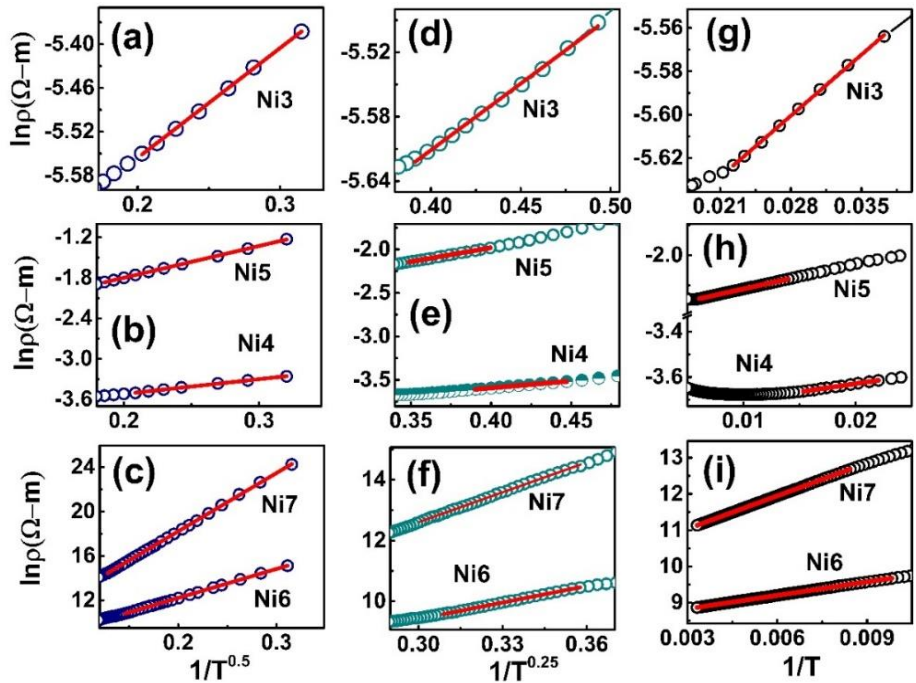

**Figure 6** Fitted resistivity data at different temperature regimes of various Ni NP samples as indicated using ES-VRH (a-c), Mott-VRH (d-f) and Arrhenius (g-i) models.

**Table 2** Fit parameters for semiconducting regimes of the Ni samples obtained from ES-VRH and Mott-VRH models and calculated ones.

| Sample | $T_{ES}$ (K) (T-range) | $E_{ES}$ (meV) | $\frac{R_{ES}}{\xi}$ | $T_M$ (K) (T-range) |
|---|---|---|---|---|
| Ni3 | 1.5   (10 K-24 K) | $0.05 \times T^{1/2}$ | $0.306 \times T^{-1/2}$ | 2.3   (17 K-42 K) |
| Ni4 | 2.8   (10 K-23 K) | $0.07 \times T^{1/2}$ | $0.418 \times T^{-1/2}$ | 6.0   (27 K-44 K) |
| Ni5 | 14.4  (10 K-30 K) | $0.16 \times T^{1/2}$ | $0.948 \times T^{-1/2}$ | 110.2  (35 K-67 K) |
| Ni6 | 671.0 (10 K-47 K) | $1.11 \times T^{1/2}$ | $6.47 \times T^{-1/2}$ | $9.17 \times 10^4$ (47 K-108 K) |
| Ni7 | 2745.0 (10 K-61 K) | $2.25 \times T^{1/2}$ | $13.09 \times T^{-1/2}$ | $1.15 \times 10^6$ (65 K-120 K) |

**Table 3** Fit parameters for semiconducting regimes of samples obtained from Mott-VRH, ES-VRH and Arrhenius models and calculated parameters.

| Sample | $E_M$ (meV) | $\frac{R_M}{\xi}$ | $T_M/T_{ES}$ | $T_{C,exp}$ (K) | $T_{C,cal}$ (K) |
|---|---|---|---|---|---|
| Ni3 | $0.30 \times T^{3/4}$ | $0.46 \times T^{-1/4}$ | 1.53 | - | - |
| Ni4 | $0.39 \times T^{3/4}$ | $0.58 \times T^{-1/4}$ | 2.14 | 27 | 21 |
| Ni5 | $0.81 \times T^{3/4}$ | $1.21 \times T^{-1/4}$ | 7.65 | 34 | 30 |
| Ni6 | $4.35 \times T^{3/4}$ | $6.52 \times T^{-1/4}$ | 136.7 | 47 | 82 |
| Ni7 | $8.18 \times T^{3/4}$ | $12.28 \times T^{-1/4}$ | 418.8 | 61 | 104 |

**Table 4** Fit parameters for semiconducting regimes of samples obtained from Mott-VRH, ES-VRH and Arrhenius models and calculated parameters.

| Sample | $\Delta_{CG}$ (meV) | $T_{CG}$ (K) | $E_C$ (meV) (T-range) | Fitted x | Calculated γ |
|---|---|---|---|---|---|
| Ni3 | - | - | 0.34 (27 K-47 K) | - | - |
| Ni4 | 0.16 | 1.91 | 0.62 (45 K-65 K) | 0.53±0.04 | 2.43±0.04 |
| Ni5 | 0.44 | 5.20 | 1.4 (64 K-166 K) | 0.54±0.03 | 2.52±0.03 |
| Ni6 | 4.92 | 57.27 | 10.7 (110 K-300 K) | 0.52±0.05 | 2.25±0.05 |
| Ni7 | 11.53 | 134.1 | 26.1 (115 K-300 K) | 0.50±0.02 | 2.0±0.02 |

There are increase in fitted and calculated parameters $T_{ES}$, $E_{ES}$, $T_M$, $E_M$, $\frac{R_{ES}}{\xi}$ and $\frac{R_M}{\xi}$ with decrease in particle size (tables 2 & 3). They correspond to extension of temperature to higher values of Mott-VRH, ES-VRH and Arrhenius models as we progress from Ni3 to Ni7. The crossover temperature $T_C$ from Mott-VRH to ES-VRH increases with decrease in particle size, which is in line with more insulating behaviour of smaller size NPs[38]. The calculated value of crossover temperature $T_{C,cal}$ is closed to experimental $T_{C,exp}$ obtained from resistivity data for Ni4 and Ni5 (table 3). It is however much higher in Ni6 and Ni7, corresponding to coexistence of both ES-VRH and Mott-VRH conduction. The increase in $T_M/T_{ES}$ and $\Delta_{CG}$ with decrease in particle size is consistent with enhanced insulating behaviour of NPs with decrease in size. However, $\Delta_{CG}$ is significantly increased in Ni6 and Ni7 and corresponds with decreased lattice parameter or coexistence of fcc and hcp phases that matches with more Coulomb interactions between electrons.

Above the ES-VRH regime, Mott-VRH well-describes the electrical conduction i.e. fixed DOS at the Fermi level may appear at high temperature. Further increase in temperature above Mott regime, activation behaviour starts to dominate. The activation energy $E_c$ obtained from Arrhenius model is found to be increased with decrease in particle size, which manifests increase in energy barrier height associated with increase in TOP coverage and hence an increase in inter-NP separation or decrease in coupling between NPs[16]. The trends in obtained parameters (tables 2- 4) are consistent with earlier reports[16,38]. It can be seen that $\frac{R_{ES}}{\xi}$ is greater than 1 below 40 K in Ni6 and below 150 K for Ni7, while $\frac{R_M}{\xi}$ is greater than 1 for Ni6 and Ni7. They are found to be unreasonably smaller than unity for Ni3 to Ni5, which is attributed to larger influence of metallic background with less semiconducting behaviour, in line with also previous report[38]. Thus, $\frac{R_{ES}}{\xi} > 1$ and $\frac{R_M}{\xi} > 1$ for Ni6 and Ni7 manifest ES-VRH and Mott-VRH conductions in their respective temperature ranges.

ES-VRH holds true for DOS at the Fermi energy with power-law dependence as[37]

$$N(E) = N_0|E - E_F|^\gamma. \qquad (12)$$

The N(E) decreases with decrease in T with $\gamma = 2.7+0.1$ for 3D material[39]. The value of $\gamma$ can be obtained using resistivity data through the exponent $x = \frac{\gamma+1}{\gamma+d+1}$, where d is the dimensionality. For Mott mechanism $\gamma=0$, where x=1/4 and for ES-VRH mechanism $\gamma=2$, where $x=1/2$. To obtain the value of x, we have followed the method given in ref[39]. Assuming the semiconducting regimes, the universal resistivity behaviour is

$$\rho = BT^{-m}\exp\left(\frac{T_0}{T}\right)^x \text{ and a parameter } w(T) = -\frac{d(ln\rho)}{dlnT} = m + x\left(\frac{T_0}{T}\right)^x \qquad (13)$$

For exponential hopping resistivity, the second term in eq. 13 is much larger than the first and hence first term (m) is considered negligible. Thus, the slope of plot between lnw versus lnT gives the value of x. The value of x and $\gamma$ obtained from linear fitting of graphs plotted between lnw versus lnT for Ni4 to Ni7 and calculated from $x = \frac{\gamma+1}{\gamma+4}$ as shown in figure 7 & table 4. The value of critical exponent x (table 4) is somewhat close to 0.5, suggesting appearance of ES-VRH conductivity due to vanishing single particle DOS at the Fermi level[37].

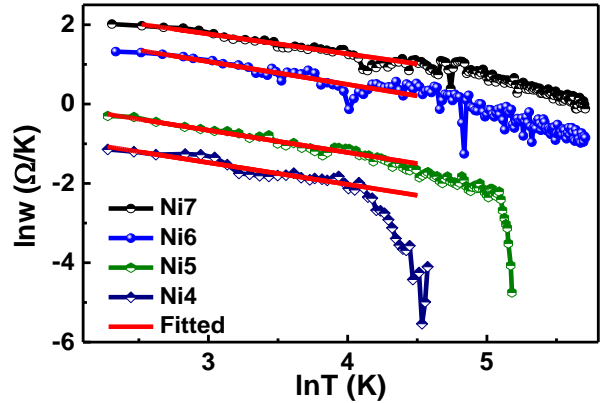

**Figure 7** Linear fitted curves of lnw versus lnT for Ni4, Ni5, Ni6 and Ni7. Red line shows the fitting.

The $\gamma$ value for Ni7 well-matches its theoretical value $\gamma=2$ for 3D, while others are slightly larger. Its value lying between about 2 to 2.5 is quite close to those of earlier reports[37,39].
At low temperature, there is increase in localized states, and charges are transported through localized states in the vicinity of the Fermi energy through hopping conduction. The sudden rise in the resistivity below Mott temperature regime is probably due to opening of $\Delta_{CG}$. Similar rise has been observed in thermopower (discussed below), which may also be due to reduced charge carrier concentration near the Fermi level because of $\Delta_{CG}$. Increase in the $\Delta_{CG}$ with decrease in particle size indicates that larger/ sharper localized DOS (eq. 10) which also gets support from increase in $T_M$ and $T_{ES}$ with decrease in particle size owing to enhanced localization of charge carriers.

Therefore, these results are mainly due to three competing phenomena as (i) increase in level spacing and band/ Coulomb gap opening, (ii) increase in TOP thickness or barrier height and (iii) decrease in NP coupling between adjacent NPs that leads to decrease in transmission or tunnelling probability of electrons[15,16,33–36]. The size (and TOP)-induced MIT is mainly due to increase in overall energy barrier i.e. (ii) and (iii), while T-driven MIT is probably owing to smaller thermal energy of electrons compared to energy barrier, wherein long-range hopping is less likely.

Significant increment in ρ and transition from metallic to semiconducting behaviour as size decreases indicates that disorder (GBs, organic/inorganic interfaces and crystal defects including point defects) increases as size decreases. In addition, surfactant used also covers the NP surface, which act as the boundary resistance by forming a metal-semiconductor-like interface. Thus, transmission coefficient of electrons is likely to decrease with decrease in particle size since TOP coverage increases. As a result, metal to semiconducting and to insulating transition behaviour of these NPs with decrease in particle size takes place. It has been noted that ρ steadily increases in 3.2 nm and 1.3 nm NPs since crystallite size is nearly equal or smaller than the bulk mean free path of electrons i.e. they are strongly confined. Understanding of such complex conduction processes may thus probably fit the caution of Beloborodov *et al.*[40] who noted that 'understanding the metal-insulator transition in granular (metallic) materials is one of the most difficult theoretical problems and considerable efforts will be necessary to solve it.'

### 3.3 Seebeck coefficient (S)

S for Ni1 to Ni7 along with annealed bulk polycrystalline Ni ingot (having purity 99.99%) as reference is shown in figure 8 (a). The S of bulk Ni decreases much slower as temperature decreases below 140 K, behaving as if it is near to saturation down to 45 K, below which S suddenly rises as temperature decreases, and shows negative sign in whole T range. The feature between 45 K to 140 K is assigned to the phonon-drag minimum (PDM). The S around -3.91 µV/K and -0.66 µV/K at 300 K for Ni1 and Ni2, respectively seen is significantly smaller than that (~ -13.5 µV/K) of the bulk Ni although, it is found to be larger with value 2.92 and 3.37 µV/K at 10 K in Ni1 and Ni2, respectively compared to its bulk value -1.61 µV/K. A crossover from positive to negative near 207 K and 275 K in S for Ni and Ni2 is seen, and this transition temperature shifts towards higher T with reduction in particle size. Interestingly, PDM is supressed in Ni1 and Ni2 compared to bulk Ni. Moreover, a broad hump-like feature near at 140 K in Ni1 and Ni2, which is shifted to higher T compared to Ni bulk that is mainly due to enhanced diffusion of electrons at sufficiently high T (figure 8 (a)). These results are consistent with earlier report[41].

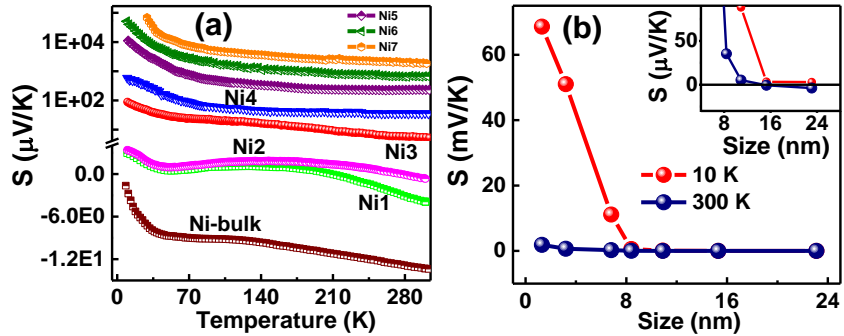

**Figure 8** (a) Temperature variation in thermopower of Ni1 to Ni7 along with bulk Ni and (b) thermopower as a function of crystallite size at 10 K and 300 K; Inset: expanded view for smaller scale.

However, these features are drastically changed with further decrease in particle size such as feeble phonon drag effect in Ni3 and Ni4, and PDM is completely supressed in Ni4 to Ni7 (figure 8 (a)). Notably, it shows positive value in whole T range in Ni3 to Ni7, and complete disappearance of PDM associated with colossal increase in its values (figure 8 (a, b)). It is clearly seen that S increases as particle size decreases in Ni1 to Ni7 at 10 K and 300 K (figure 8 (b)), and they show the change in conduction from n-type to p-type (figure 8 (b), inset). These NPs thus exhibit colossal positive S in Ni4 to Ni7 at 10 K and very large values even at 300 K. It may be pointed out that S could not be measured due to larger resistance of Ni7 below 30 K (figure 4 (a)) than the input impedance ($10^{10}\,\Omega$) of nanovoltmeter used to measure S.

Now, the relation between change in band structure and Seebeck coefficient for single band conduction degenerate semiconductor can be understood from eq. 14.

$$S = -\frac{\pi^2 k_B^2 T}{3e}\left[\frac{dn}{ndE} + \frac{1}{\mu}\frac{d\mu}{dE}\right]_{E=E_F} \qquad (14)$$

S for metals and degenerate semiconductors with parabolic band in terms of effective mass ($m^*$) and carrier concentration ($n$) can be expressed by Mott's relation[1,4]

$$S_d = -\frac{8\pi^2 k_B^2 T}{3eh^2} m^* \left(\frac{\pi}{3n}\right)^{2/3}. \qquad (15)$$

where $n(E)$ (is the function of localized DOS) and $\mu(E)$ are energy-dependent carrier concentration and mobility, respectively, and $m^*$ is effective mass of charge carrier. Therefore, it is clear from eq. 14 that larger the DOS at the $E_F$, the stronger the energy dependence on $\mu(E)$, which enhances S. The overall electrical transport behaviour is in fact highly dependent on crystallite size, size distribution, shape and surface-bound molecules due to spatial confinement of electrons, phonons[42] and their additional scattering with impurities (defects, GBs, surfaces and surfactants as a consequence of the overall increase in relaxation rates and electron-phonon scattering[42].

From eq. 14, it is clear that S will enhance mainly in two ways, first increase in dn(E)/dE i. e. local increase in DOS that is associated with QSE, and second increase in $\mu(E)$ i. e. a strong energy dependence of relaxation time $\tau(E)$ or scattering mechanism that strongly depend on the energy of the charge carriers. As TOP increases, particle size decreases and NPs become monodispersed, which may introduce sharp rise in DOS and hence S is likely to increase. Furthermore, enhanced DOS(E) near Fermi level makes stronger energy dependence of $\mu(E)$ that also support the higher absolute value of S with decrease in crystallite size. The magnitude of S depends upon derivative of energy-dependent conductivity $\sigma(E)$ or electron lifetime ($\tau(E)$)[31,42] and hence negative to positive crossover with decrease in particle size may take place. Moreover, the parabolic band approximation may no longer be valid in these NPs due to distortion in Fermi surface and then sign and magnitude will also depend on $m^*$. This may give negative or positive value of $m^*$ and accordingly sign of S (eq. 15). Since they are semiconductors combined with semiconducting smaller NPs, they will lead to colossal resistivity and S as size decreases. Further, the increase in ρ with decrease in particle size is more likely due to scattering of charge carriers, which suggest that strong energy dependence on $\tau(E)$. As a result, as particle size decreases, decrement in $n$ (eq. 14) and enhancement of DOS lead to enhancement in S. Therefore, the colossal value of S would also contribute from the first term in eq. 14 i.e. associated with local enhancement of DOS, as observed earlier[21].

### 3.4 Thermal Conductivity (κ)

The temperature evolution of κ for Ni1, Ni3, Ni4, Ni5 and Ni7 is shown in figure 9. κ of Ni1 gradually increases as T increases associated with a broad hump-like feature around 200 K. Such trend has been nearly changed in Ni3 to a peak near 32 K with the broad hump-like feature around 200 K still persisting. These features are respectively increasing and decreasing, and finally, turn completely linear as one goes from Ni3 to Ni7 (see figure 9, inset (a) for clarity). That is, peak becomes more distinct, the clearest and sharpest being in Ni7 while hump-like feature slowly turns to straight one as the size decreases. The value of κ at 300 K falls fast nearly linearly but that at 10 K increases slowly with decrease in particle size (figure 9, inset (b)). The peak values around 0.72 W/m-K at 32 K, 0.70 W/m-K at 32 K, 0.58 W/m-K at 27 K and 0.49 W/m-K at 28 K are found for Ni3, Ni4, Ni5 and Ni7, respectively. There are thus trends of evolution of decreasing κ values, changing shape and slope with decrease in particle size while the peak-like structure remains in the range 27 K - 32 K. They are ascribed to QSE, modified DOS, Coulomb gap and metal/organic interface plus the accompanying nanolattice of these NPs[26]. In overall, thermal conductivity drops quite significantly with decrease in particle size, which is consistent with those of nanowire of Ni and Ag[43,44]. The observed κ ~ 0.52 W/m-K at 300 K in Ni7 is ultralow and around 1/175 of Ni bulk of 91.0 W/m-K[45]. This is comparable with earlier reported values in complex TE material $TiS_2$[3], $(SnTe)_{1-2x}(SnSe)_x(SnS)_x$[10], Au nanocrystal arrays and other materials[14]. Most importantly, this value is significantly smaller than the well-known other TE materials like $Bi_2Te_3$, bulk silicon and nanocrystalline silicon[46], showing probable potential in thermoelectrics, heat dissipation in nanocrystal arrays-based electronics and photonics[14]. The lowest value of κ is ~ 0.13 W/m-K at 10 K in Ni7.

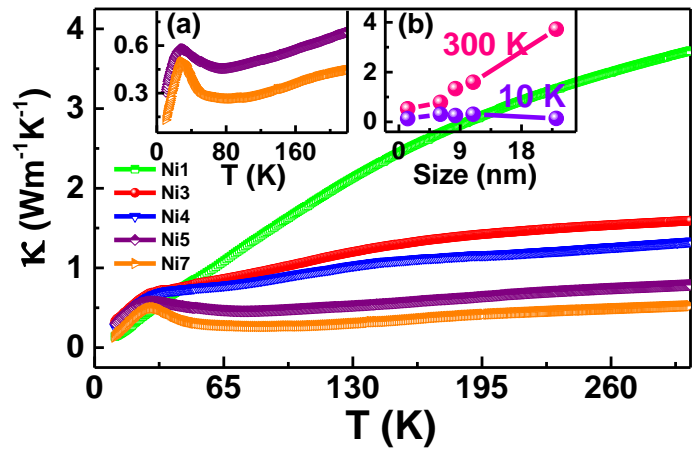

**Figure 9** Thermal conductivity versus temperature for Ni1, Ni3, Ni4, Ni5 and Ni7. Inset: (a) expanded view of Ni5 and Ni7 and (b) thermal conductivity at 10 K and 300 K as a function of size.

As we have seen in electrical resistivity and Seebeck coefficients, these NPs become gradually more towards insulating nature from metallic state showing their very intriguing behaviours as size decreases. In metals, electrons usually transport thermal energy grossly while phonons do it in smaller fraction of it[47]. The ratio of thermal conductivity $κ_e$ to electrical conductivity is proportional to temperature due to electronic part in a crystalline metal (Wiedemann–Franz law). As a result, $κ_e$ is constant at T > $θ_D$, inversely proportional to $T^2$ at T < $θ_D$ and proportional to T at T << $θ_D$. This leads to a peak between $T^{-2}$ and T dependences before the impurity scattering takes over. In a crystalline insulating material, since the kinetic formula for thermal conductivity κ is proportional to the product of total heat capacity, average speed and mean free path, κ is inversely proportional to T for T > $θ_D$, exponentially T-dependent (due to U-processes) and $T^3$-dependent (in the boundary scattering region) at T < $θ_D$. Between U-processes and boundary scattering region, a peak appears. Such features are completely missing in Ni1 showing the dominant role of GBs, lattice imperfections, missing lattice points and surfactants, which respond differently to electrons and phonons. Moreover, Ni1 contains randomly sized NPs. This however is not so in other samples, wherein NPs are (more) uniform and hence form their own lattice in addition to the atomic lattices[26]. Nanolattice along with increasingly enhanced disorders (GBs, lattice imperfections, point defects, surfactants), modified DOS, QSE and metal/organic interface is considered to be responsible for the gradual enhancement of the peak feature near 30 K associated with gradual drop in κ values as the crystallite size decreases.

### 3.5 Power factor ($S^2\sigma$) and figure of merit (ZT)

Now, power factor as a function of T is shown in figure 10 (a). A very sharp dip near 207 K in Ni1 with slight rise below 50 K, above which they remain gradually dropped or rise with temperature. Dip-like structures are correlated with the crossover of S from negative values to positive values near these temperatures and hence do not arise in remaining samples. Additionally, for Ni7, there is slight saturation and then a drop in these values at around 35 K and these values also are lowest amongst them. $S^2\sigma$ at 10 K and 300 K increases as particle size decreases from Ni1 to Ni3, and then decreases in Ni4 and Ni5 with the smallest or largest in Ni7 (figure 10 (a), inset). The highest value of $S^2\sigma$ is 537.10 μW/m-K$^2$ at 10 K in Ni5 and 2.95 μW/m-K$^2$ in Ni1 at 300 K.

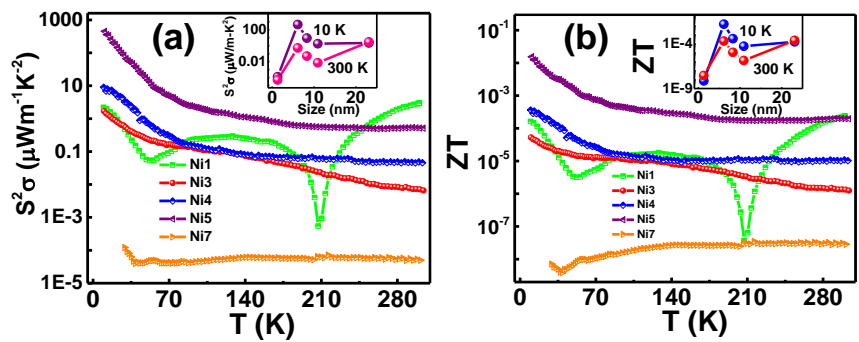

**Figure 10** (a) Power factor $S^2\sigma$ and (b) figure of merit ZT of Ni1, Ni3, Ni4, Ni5 and Ni7; Insets: variation of these parameters at 10 K and 300 K as a function of particle size.

The temperature-dependence on ZT (figure 10 (b)) looks to be similar to $S^2\sigma$ (figure 10 (a)), and similar trends at 10 K and 300 K are seen in size dependence (figure 10 (b), inset). The highest ZT is ~ 0.02 in Ni5 at 30 K and 2.38×10$^{-4}$ in Ni1 at 300 K (figure 10 (b), inset). Moreover, when size is below 5 nm, both these parameters are drastically affected. The reason behind these trends is closely related to simultaneous drop (rise) in electrical conductivity (resistivity; due to insulating organic ligands) and thermal conductivity. As a result, ZT could not be improved to a very significant value even though these NPs possess a very rich and complex features like QSE, modified DOS, metal/organic interface and nanolattice. The significantly high-power factor and considerable enhancement in ZT at 10 K in Ni5 along with its low κ ~ 0.30 W/m-K observed are much smaller than the well-known TE material CsBi$_4$Te$_6$ at this temperature[5]. Thus, these results will provide a vista for future search for novel thermoelectric materials.

### Conclusion

We demonstrate here size and organic ligand/s mediated metal to insulator transition, colossal Seebeck coefficient, ultralow thermal 0.52 W/m-K at 300 K and enhanced figure of merit in solution-processed monodispersed nickel NPs. In addition, large negative temperature coefficient of resistance and n-type conduction to p-type conduction with decrease in particles are observed. These results are interpreted in terms of formation of metal/organic interfaces, sharp increase in density of states and multiscale electron and phonon scattering by various defects. Thus, this study will open a new avenue to make better thermoelectrics through incorporation of such NPs in semiconducting hosts.

### Acknowledgements

Authors would like to acknowledge Dr. D. M. Phase and Dr. M. Gupta, UGC-DAE Consortium for Scientific Research, Indore, India for providing XRD data. TEM data were obtained from MRC, MNIT, Jaipur by self-funding.

### Authors contribution

VS planned the experiments, prepared the samples and performed the thermopower and resistivity measurements, and YKK provides the thermal conductivity data. VS did the analysis with help from GSO and YKK, and VS wrote the manuscript and GSO corrected and finalized the paper